\documentclass[conference]{IEEEtran}
\IEEEoverridecommandlockouts

\usepackage{cite}
\usepackage{amsmath,amssymb}
\usepackage{graphicx}
\usepackage{booktabs}
\usepackage{multirow}
\usepackage{url}
\usepackage{stfloats}
\usepackage[hidelinks]{hyperref}
\usepackage{xcolor}
\usepackage{verbatim}
\usepackage{cite}

\graphicspath{{figures/}}

\begin{document}

\title{AMORT-V1: Checkpointed Post-Quantum Attestation for Industrial Controllers}


\author{
\IEEEauthorblockN{
Rishabh Das\textsuperscript{1},
Gerald Agendia\textsuperscript{1},
Jibriel Adjei\textsuperscript{1},
Ahmed Oun\textsuperscript{2}
}
\IEEEauthorblockA{
\textsuperscript{1}J. Warren McClure School of Emerging Communication Technologies, Scripps College of Communication\\
\textsuperscript{2}School of Electrical Engineering and Computer Science, Russ College of Engineering and Technology\\
Ohio University, Athens, OH, USA\\
\{rishabh.das,ga097624,ja501625,oun\}@ohio.edu
}\vspace{-1.1cm}
}

\maketitle

\begin{abstract}
Programmable logic controllers often remain deployed for decades, making post-quantum attestation a long-term requirement. Prior measurements of OpenPLC attestation show that post-quantum signing can consume 88--99\,\% of round-trip latency and that durable state can further limit throughput. We introduce AMORT-V1, a checkpointed attestation protocol that moves public-key signing and durable commitment from individual reports to periodic checkpoints. An ML-KEM epoch establishes per-device evidence keys, devices authenticate reports with MACs, and the controller commits ordered evidence through hash-chained Merkle checkpoints signed with ML-DSA or SLH-DSA. Because shared MACs do not provide transferable proof of device origin, AMORT-V1 adds a device-held disclosure chain whose keys are released after checkpoint commitment. We validate the protocol with a 27-case attack matrix, 149 differential-test vectors, and a seven-node Raspberry Pi~5 testbed with five physical OpenPLC provers. Checkpointing reduces controller cost by $90\times$ relative to per-report ML-DSA without durable logging and by $878\times$ when durable logging is required. Transferable device-origin verification costs 2.95\,$\mu$s per report, while controller-side scaling to 10\,000 identities requires 6.18\,$\mu$s and 155 octets per report. AMORT-V1 also supports rollback-resistant restart through TPM-backed monotonic state without placing durable hardware operations on the per-report path.
\end{abstract}


\section{Introduction}

Shor's algorithm allows a quantum computer to forge RSA and elliptic-curve signatures~\cite{shor}. Programmable logic controllers (PLCs) are exposed to that forgery risk for longer than most systems, because operators keep them in service for decades. A signature scheme chosen for a PLC today will still be in the field when quantum computers become practical~\cite{cnsa2}. The signatures matter because they protect the controller's software. An attacker who alters that software can tamper the physical process and hide the alteration, as Stuxnet did when it reprogrammed centrifuge logic~\cite{langner}. Remote attestation defends against such alterations by letting a verifier check the software state of a PLC. The check depends on signatures over the reported evidence. If a quantum adversary can forge those signatures, the verifier cannot detect the tampering. Post-quantum attestation must therefore protect the controller's software state, and the cryptographic evidence used to establish that state.

In an industrial deployment, attestation must extend beyond a single controller because many controllers may report their state repeatedly to a common verifier. Prior work addresses the first part of this scaling problem. SEDA and SANA reduce verifier workload by aggregating evidence from thousands of devices within an attestation round~\cite{seda,sana}, while Liu et al.\ demonstrate that low-end provers can support post-quantum attestation~\cite{liu}. These results show that attestation can scale across devices and that post-quantum mechanisms can run on constrained platforms. They do not, however, reduce the cryptographic work that recurs each time a controller produces new attestation evidence. When many controllers attest repeatedly, per-report signing and state-management costs accumulate across both devices and time.

A prior study measured that per-report work for post-quantum attestation on OpenPLC~\cite{openplc1}. Hash-based signing consumed 88-99\,\% of each attestation round trip. Measurement, encoding, decoding, and policy evaluation together took microseconds. Network transport added latency on links slower than 10\,Mb/s. Stateful hash-based signatures added a second bottleneck: the signer had to persist its one-time-key state before releasing each signature, and that write limited the system to 217 synchronous operations per second. Both bottlenecks recur with every report. A controller reporting at 1Hz generates one signature per second, while a stateful signer also performs one durable write per second, even when the measured software state remains unchanged. The total cost of attestation therefore grows with both the number of controllers and their reporting rate. Existing aggregation techniques reduce verifier-side cost as the controller population grows, but they do not reduce the recurring cost incurred by each controller.

To address this scaling problem, we introduce AMORT-V1, which authenticates individual attestation reports with symmetric evidence and performs asymmetric authentication and durable commitment only at periodic checkpoints. This design reduces how often the system performs signing and durable-state operations, and changes the security properties attached to each report. A signed checkpoint proves the order of reports and binds them to an append-only history. A MAC does not prove which device produced a report, because the controller verifying the MAC holds the same key and could have produced the report itself. The design question is therefore twofold: which security properties survive when signatures move to checkpoint boundaries, and how the protocol recovers the property that a MAC cannot provide. We study three research questions.

\begin{itemize}
\item[\textbf{RQ1}] How does the checkpoint interval trade amortized post-quantum cost against omission-detection delay (\S\ref{sec:model})?
\item[\textbf{RQ2}] Can AMORT-V1 recover publicly transferable device-origin authentication after replacing per-report signatures with symmetric evidence, and at what cost (\S\ref{sec:audit}, \S\ref{sec:twotag})?
\item[\textbf{RQ3}] Can per-device epochs preserve failure isolation and support recovery from revocation, re-keying, partition, and prover restart without reintroducing per-report durable state (\S\ref{sec:recovery})?
\end{itemize}

This work makes three contributions.

\begin{itemize}
\item \textbf{Checkpointed post-quantum attestation.} AMORT-V1 derives a per-device evidence key from an ML-KEM epoch, authenticates each report with a MAC under that key, and commits reports through hash-chained Merkle checkpoints signed with ML-DSA or SLH-DSA.

\item \textbf{Transferable device-origin authentication under symmetric evidence.} Checkpointing preserves ordering and exposes equivocation, but the controller holds the MAC key and so a MAC cannot establish authorship. AMORT-V1 adds a disclosure chain that only the device can compute, which lets a third party verify authorship after commitment. We measure the delay and verification cost of that chain.

\item \textbf{Evaluation across cost, scale, and recovery.} We validate the protocol against a 27-case attack matrix and 149 differential test vectors, then evaluate it on a seven-node Raspberry~Pi~5 testbed with five physical OpenPLC provers. The evaluation covers the checkpoint cost--detection tradeoff, controller scaling to 10\,000 identities, execution within a 10\,ms control cycle, and restart and rollback recovery.
\end{itemize}

\section{Background and Related Work}

Post-quantum signature schemes impose different costs on repeated attestation. ML-DSA~\cite{fips204} provides fast lattice-based signing, while SLH-DSA~\cite{fips205} uses stateless hash-based signatures with higher signing and communication costs. Stateful LMS and XMSS~\cite{sp800208} require the signer to prevent one-time-key reuse and rollback, making durable state part of their operation~\cite{hbs-state}. These differences matter for attestation because controllers generate evidence repeatedly, turning signature latency and durable-state operations into recurring costs.

Previous attestation systems focus on reducing verifier workload as the number of devices grows. SEDA~\cite{seda} organizes devices into an attestation structure and aggregates their results so that a verifier does not process one independent report from every device. SANA~\cite{sana} similarly uses aggregation to support attestation across large device populations, while Liu et al.\ show that low-end provers can support post-quantum attestation~\cite{liu}. Aggregation reduces the number of values a verifier processes within an attestation round, whereas AMORT-V1 reduces how often the system performs public-key signing and durable commitment across successive reports. The two approaches are therefore complementary: one scales across devices, while the other amortizes cost across time.

Moving public-key operations to checkpoints requires a mechanism that commits reports to an ordered history. Certificate Transparency provides this foundation through append-only Merkle logs, inclusion proofs, consistency proofs, and signed tree heads~\cite{rfc6962,rfc9162}. Its gossip model also shows why a signed log cannot prevent a log operator from presenting different histories to isolated verifiers. AMORT-V1 adopts the RFC~6962 hashing construction and uses hash-linked signed checkpoints to commit attestation records. A log, however, establishes only that a record appears in a committed history, it does not establish which device produced that record.

Recovering device origin therefore requires additional workflow beyond checkpoints. TESLA creates asymmetric verification properties from symmetric cryptography by having a sender commit to a one-way key chain, authenticate messages with keys that remain secret when the messages arrive, and disclose those keys later~\cite{tesla}. AMORT-V1 adapts this principle to attestation. Instead of releasing keys according to synchronized time intervals, a device releases its disclosure key only after a signed checkpoint commits the corresponding report. The checkpoint therefore replaces elapsed time as the disclosure condition, removing the need for TESLA-style clock synchronization while tying public device-origin verification to the checkpoint interval.

The need for authenticated software state is well established in industrial control systems. Basnight et al.\ demonstrated firmware-modification attacks against PLCs~\cite{basnight}, while Yoo and Ahmed demonstrated control-logic injection attacks~\cite{yoo}. Such attacks can change the logic that governs a physical process while leaving the controller operational. OpenPLC provides an IEC~61131-3-compatible platform for studying these attacks and related defenses~\cite{openplc}, and Alsabbagh et al.\ identify authentication and replay weaknesses in the platform~\cite{alsabbagh}. These studies provide the foundations for scalable attestation, but do not address repeated post-quantum evidence generation over time. This work introduces AMORT-V1 protocol, which combines these ideas to reduce per-report post-quantum cost without giving up ordered commitment, omission detection, or transferable device-origin authentication.

\begin{figure*}[t]
\centering
\includegraphics[width=\textwidth]{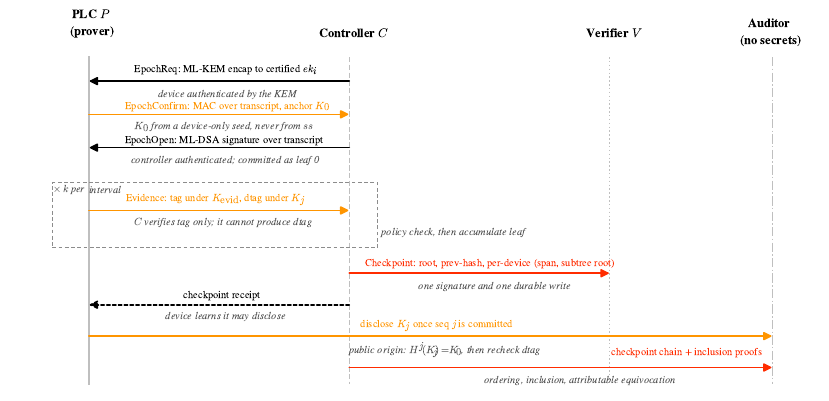}
\caption{AMORT-V1. One ML-KEM exchange and one controller signature open an epoch. Each of the next $k$ reports carries a controller-verifiable tag and, for public-origin verification, a second device-only disclosure tag. One signed checkpoint commits the batch, after which the device releases the disclosure key for third-party origin verification. Orange marks the symmetric evidence path; red marks the checkpoint path.}
\label{fig:protocol}
\vspace{-.3cm}
\end{figure*}

\section{Threat Model}\label{sec:threat}

We evaluate AMORT-V1 against three adversary classes and a cold-auditor verification condition.
\begin{itemize}
    \item A \emph{network adversary} can observe, inject, replay, reorder, delay, truncate, and drop attestation traffic, but cannot break the underlying cryptographic primitives. 
    \item A \emph{compromised device} that can control its own software state and epoch secrets from the point of compromise onward and can therefore submit arbitrary measurements or malformed protocol messages under its own identity.
    \item A \emph{malicious controller} that can suppress records, alter their commitment order, or present conflicting checkpoint histories to different verifiers. This adversary is central to AMORT-V1 because the controller also holds the symmetric evidence key and can reproduce ordinary report MACs.

\end{itemize}

Under the \emph{cold-auditor verification condition}, the auditor does not hold the
secret key material and the trusted local history, but must still verify the integrity
and ordering of checkpoints, as well as any transferable device-origin evidence.

The threat model excludes physical key extraction, compromise of the trusted measurement agent, runtime-memory attacks that occur after measurement, and denial of service through flooding or resource exhaustion. AMORT-V1 authenticates the freshness, origin, and committed history of a reported software-state claim, but it does not establish hardware-rooted integrity of the measurement agent or guarantee that the reported state itself is correct.

\section{AMORT-V1}\label{sec:protocol}

AMORT-V1 moves asymmetric authentication and durable commitment from individual reports to periodic checkpoints and preserves the security properties needed for attestation. The design targets six properties: \emph{Freshness} limits the age of accepted evidence. \emph{Device binding} associates an accepted report with a device identity and the epoch in which it was generated. \emph{Omission detection} exposes suppressed evidence within a bounded interval. \emph{Public verifiability} allows a party holding no secret to verify checkpoint inclusion and ordering. \emph{Containment} limits the effect of a device compromise to that device and its current epoch. Finally, \emph{detectable equivocation} allows conflicting checkpoint histories to serve as evidence of controller misbehavior once the conflicting views are compared.

The protocol has five stages, and this section follows a report through them (Figure~\ref{fig:protocol}). An epoch begins with an ML-KEM exchange that authenticates the device and the controller and yields the epoch's keys. The device then sends reports, each carrying one or two MAC tags. The controller verifies each report and adds it to a Merkle tree. At the checkpoint, the controller signs the tree root together with a sequence range and a subtree root for each device. After the checkpoint, the device releases a disclosure key so that a third party can verify authorship. A final stage handles restart and rollback.

\subsection{Epoch Establishment and Key Derivation}
Device $i$ holds a static ML-KEM-768 decapsulation key, and a certificate binds the matching public key to $\mathrm{dev\_id}$. The controller opens an epoch by encapsulating to that public key. Only device $i$ can decapsulate the result to obtain the shared secret $ss$, and the device proves that it did so by returning a MAC over the exchange transcript. Both sides then derive two keys from $ss$:
\begin{align*}
K_{\mathrm{ep}}
  &= \mathrm{HKDF\text{-}Extract}
     \bigl(H(\mathrm{transcript}), ss\bigr),\\
K_{\mathrm{evid},i}
  &= \mathrm{HKDF\text{-}Expand}
     \bigl(K_{\mathrm{ep}}, \mathrm{info}_i\bigr),
\end{align*}
where
\[
\mathrm{info}_i =
\texttt{"evidence"}\,\|\,\mathrm{dev\_id}\,\|\,\mathrm{epoch}.
\]
The evidence key $K_{\mathrm{evid},i}$ authenticates every report in the epoch. One ML-KEM exchange per device per epoch therefore replaces one signature per report.

The exchange authenticates the device but not the controller. Only device $i$ can decapsulate, so a completed confirmation proves the device's identity. Any party that knows the device's public key can encapsulate to it, so a completed exchange proves nothing about the controller. An unauthenticated controller could open a rogue epoch, force the device to replace its keys, or advance its sequence state, and each of those actions disrupts attestation.

The controller therefore signs an \textsc{EpochOpen} record with its long-term ML-DSA-65 key. The record contains the controller and device identifiers, the epoch number, the tag length, the policy identifier, the disclosure-chain anchor, an expiry, the CA certificate serial, and the transcript hash. The device accepts the epoch only if the signature verifies, the epoch number exceeds the device's current epoch, and the transcript hash matches the exchange the device observed. After \textsc{EpochOpen}, both parties are authenticated: ML-KEM authenticated the device, the ML-DSA signature authenticated the controller, and the transcript hash ties both to the same parameters. A deployment that authenticates the controller through a provisioning channel may omit the signature, but it must state that assumption.

One key must remain outside the shared secret. The controller knows $ss$, so it can compute every value derived from $ss$. If the disclosure chain of \S\ref{sec:audit} were derived from $ss$, the controller could compute every disclosure key, forge a report, and tag it as the device would have. The device therefore draws the disclosure seed from its own entropy, keeps the seed, and sends the controller only the chain anchor $K_0$, which the controller places in \textsc{EpochOpen}. The seed never leaves the device, and that separation is what allows a third party to trust a disclosure tag.

\subsection{Per-Report Evidence Authentication}
Every report follows the same canonical encoding, with length-prefixed fields under a domain tag. The record occupies 83 octets and carries the protocol version, the device identifier, the epoch, a per-device sequence number, a monotonic timestamp, the measurement digest, the policy identifier, and flags. Each record ends with a truncated HMAC-SHA-256 tag. The protocol supports tag lengths of 8, 12, or 16 octets and uses 16 octets by default. \textsc{EpochOpen} fixes the tag length for the epoch, and the verifier rejects any record that uses a different length.

The tag length determines the probability that an adversary can forge a report over the lifetime of the deployment. A tag of $\ell$ octets yields an online forgery probability of $2^{-8\ell}$ per attempt. A deployment of 10\,000 controllers reporting at 1\,Hz for twenty years verifies $6.3\times10^{12}$ tags. An adversary who submits forgeries at the same rate succeeds with probability near $3.4\times10^{-7}$ against 64-bit tags and $1.9\times10^{-26}$ against 128-bit tags. We report 64-bit results as an experimental point and recommend 128-bit tags for deployment.

A report can carry two tags with different purposes. The \emph{evidence tag} uses $K_{\mathrm{evid},i}$. The controller verifies it on arrival and rejects a failing report before any policy evaluation. The \emph{disclosure tag} uses a key $K_j$ from the device-only disclosure chain. Before disclosure, only the device knows $K_j$, so the controller and third parties cannot verify this tag. The evidence tag therefore supports immediate verification by the controller, while the disclosure tag supports later device-origin verification after the device releases $K_j$. Using both adds one extra MAC computation and one extra tag per report.

Every report also carries a per-device sequence number that increases by one with each report. Each checkpoint records the sequence range committed for each device. A gap between consecutive ranges reveals a missing report. Sequence ranges alone cannot detect substitution, because the controller could drop one report, insert another, and still preserve a continuous range. The next subsection adds a per-device commitment to detect this case.

\subsection{Checkpoint Construction and Ordered Commitment}
The controller adds each verified report to a Merkle tree using the RFC~6962 hashing rules. At each checkpoint, it signs a record containing the tree root, the hash of the previous checkpoint, the total leaf count, and one tuple $(\mathrm{dev\_id}, \mathrm{seq}_{lo}, \mathrm{seq}_{hi}, \mathrm{subtree})$. for each device in the batch. The $\mathrm{subtree}$ value is the Merkle root computed over that device's reports.

The previous-checkpoint hash chains the checkpoints into one history. If the controller signs two different checkpoints for the same position, the conflicting signed records provide evidence of equivocation.

The per-device subtree root detects substitution. The global Merkle root commits to all leaves, but it does not show which leaves belong to each device. A controller could therefore drop report $(A,3)$, insert another leaf, preserve the total leaf count, and still declare the range $A\!:\!1..5$. Both the leaf-count and range-continuity checks would pass. Attack case A20 exposed this weakness in an earlier protocol revision that committed only the sequence range. AMORT-V1 closes the gap with a per-device subtree root. A verifier that holds device $A$'s reports can recompute the subtree root and compare it with the committed value.

The range and leaf-count checks use only checkpoint metadata, while the subtree check requires the underlying device reports. The range and leaf-count checks use only the checkpoint chain, so a cold auditor can run them to detect declared gaps, missing device commitments, and inconsistencies between sequence ranges and tree growth. The subtree check requires the underlying device reports. Only the device or a verifier that received those reports can recompute the subtree root. The checkpoint chain alone therefore cannot detect every form of omission.

Checkpointing also reduces durable-state cost. Instead of writing once per report, the controller performs one durable write per checkpoint to commit the entire batch. Under ML-DSA, the evaluation shows that amortizing these writes provides a larger benefit than amortizing the signatures.

\subsection{Transferable Device-Origin Authentication}\label{sec:audit}

AMORT-V1 therefore supports three mechanisms that provide different levels of public origin verification.

\textbf{(a) Log only.} An auditor verifies checkpoint signatures, hash-chain links, inclusion proofs, and sequence continuity. Together, these checks establish ordering, inclusion, and detectable equivocation. They do not establish device origin because the controller also knows the evidence MAC key. The auditor must therefore trust the controller when attributing each record to a device.

\textbf{(b) Disclosure chain.} The device derives a one-way chain of length $L$ from its device-only seed, $K_{j-1}=H(\texttt{dom}\,\|\,K_j)$, and commits the anchor $K_0$ in the signed \textsc{EpochOpen}. It tags report $j$ under $K_j$ and withholds $K_j$ until a signed checkpoint commits report $j$. After disclosure, the auditor hashes $K_j$ back to $K_0$ to confirm that the key belongs to the chain, then recomputes the tag. The controller cannot forge a disclosure tag before commitment, because it does not know any undisclosed $K_j$. It cannot substitute a record after commitment, because the signed root already fixes the record. The chain costs one MAC and one tag per report, and the auditor waits at most one checkpoint interval before it can verify authorship.

\textbf{(c) Device sub-checkpoint.} The device accumulates its own $k$ reports, builds a Merkle root over them, and signs the root with its ML-DSA key. The signature proves authorship immediately and without the controller's involvement. It also reintroduces a post-quantum signature at the device, amortized over $k$ reports.

\subsection{Device Restart and State Recovery}
Checkpointing removes the controller’s per-report durable write, but the device still maintains state that must remain consistent across restart. This state includes the current sequence number, the disclosure-chain position, and the highest disclosed key. If the device restarts from stale state, it may reuse a sequence number and break omission detection, or reuse a disclosed key and allow another party to authenticate a report as the device.

AMORT-V1 provides two recovery paths. Under \emph{fresh-epoch recovery}, a restarted device discards its current disclosure state and opens a new epoch. Any reports generated after the last checkpoint remain uncommitted and are lost with the old epoch, so the protocol must make that loss visible. The controller therefore commits an \textsc{EpochClose} record containing
$\{\mathrm{dev\_id}, \mathrm{epoch}, \mathrm{last\_committed\_seq}, \mathrm{last\_seen\_seq}\}$.
If the last committed and last seen sequence numbers differ, the verifier raises \texttt{ERR\_EVIDENCE\_INCOMPLETE}. A verifier may also raise \texttt{ERR\_RESTART\_RATE} when a device restarts too frequently.

Under \emph{rollback-protected recovery}, the device anchors its sequence state to a monotonic hardware counter and resumes the current epoch after restart. The counter prevents reuse of sequence numbers and previously disclosed key state. This preserves epoch continuity without requiring a durable write for every report. Section~\ref{sec:recovery} implements the anchor with a discrete TPM~2.0 and measures its recovery cost.

\section{Evaluation}\label{sec:eval}

\subsection{Implementation and Validation}
We implemented AMORT-V1 in two forms to separate protocol correctness from performance measurement. First, we built a Python reference model that defines message encodings and protocol state transitions for reporting, checkpointing, disclosure, and error handling. The reference model lets us validate protocol behavior before introducing implementation-level optimizations. We tested it against 27 adversarial cases covering cryptographic tampering, protocol-semantic violations, and malformed encodings (Appendix~\ref{app:attacks}). All 27 cases produced the expected rejection code, while five positive controls accepted valid protocol behavior.

We then built a C implementation of the performance-critical path using liboqs for ML-KEM and checkpoint signatures and OpenSSL for SHA-256 and HMAC. We used C for timing experiments so that interpreter overhead did not distort the measured protocol cost. Before collecting measurements, we ran 149 differential test vectors through both implementations to confirm that the optimized implementation preserved the behavior of the reference model. Both produced identical outputs and state transitions.

\subsection{Testbed}
We use seven Raspberry~Pi~5 nodes. Node $P$ has four Cortex-A76 cores and 8GB of memory, runs Debian~12 with OpenSSL~3.0 and liboqs, and hosts OpenPLC and the control-cycle harness. Node $C$ also has 8GB of memory, runs Debian~13 with OpenSSL~3.5.6 and liboqs, and serves as the controller. Five additional nodes, $F_1\ldots F_5$, each have four cores and 2GB of memory and run Debian~13, OpenSSL~3.5.6, the same liboqs commit, and OpenPLC~v3 \cite{predecessor, openplc,openplc1,openplc2}. We use $F_1\ldots F_5$ for the full-image and recovery experiments in \S\ref{sec:image} and \S\ref{sec:recovery}. The remaining experiments use $P$ and $C$.

Thermal behavior determines the CPU settings. Without active cooling, $P$ and $C$ idle near 74,\textdegree C and throttle under sustained load at 2400,MHz. We therefore lock $P$ to 1500MHz for single-node and real-time experiments and run $C$ at 2400MHz for controller-scaling experiments. Active cooling keeps $F_1\ldots F_5$ between 47 and 52,\textdegree C, below the throttling threshold. For each run, we record the code version, CPU affinity, clock frequency, temperature, and throttle state, and we discard any sample collected while throttling is active.

\subsection{Cost Characterization}\label{sec:constants}

The checkpoint model in \S\ref{sec:model} depends on three cost classes: per-report work, per-epoch key establishment, and per-checkpoint signing and persistence. This experiment measures each protocol operation on the controller and groups the operations by how often the protocol runs them.

We ran five trials per configuration on $C$ at 2400\,MHz, opening a fresh epoch with fresh keys for each trial. Within a trial, we repeated each operation 1\,000 times; SLH-DSA signing and the storage probe used 200 repetitions. We report medians, p95, and p99 with percentile bootstrap intervals from 10\,000 resamples, compare paired conditions with the Mann--Whitney $U$ test and Cliff's $\delta$, and keep the raw samples.

\begin{figure}[t]
\centering
\includegraphics[width=\columnwidth]{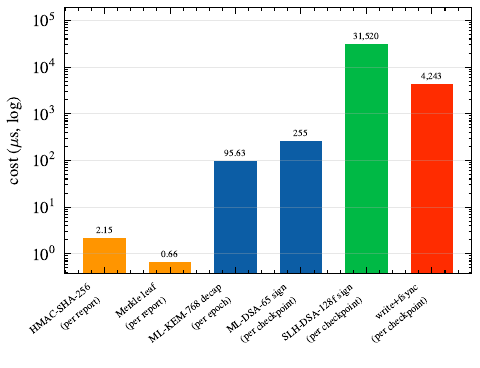}
\caption{Measured execution time of AMORT-V1 operations, grouped as per-report, per-epoch, and per-checkpoint costs}
\label{fig:constants}
\end{figure}

\begin{table}[t]
\caption{Measured protocol-operation costs on controller $C$ at 2400\,MHz.}
\label{tab:constants}
\centering
\begin{tabular}{@{}llrr@{}}
\toprule
Operation & Frequency & Median & p99 \\
\midrule
HMAC-SHA-256, 83-octet record & per report      & 2.15\,$\mu$s & 2.17\,$\mu$s \\
Merkle leaf and interior hash & per report      & 0.66\,$\mu$s & 0.67\,$\mu$s \\
\midrule
ML-KEM-768 encapsulate        & per epoch       & 63.0\,$\mu$s & 65.5\,$\mu$s \\
ML-KEM-768 decapsulate        & per epoch       & 95.6\,$\mu$s & 97.9\,$\mu$s \\
\midrule
ML-DSA-65 sign                & per checkpoint  & 254.9\,$\mu$s & 860\,$\mu$s \\
SLH-DSA-128f sign             & per checkpoint  & 31.52\,ms & 31.60\,ms \\
4\,KiB \texttt{write}+\texttt{fsync} & per checkpoint & 4.24\,ms & 7.37\,ms \\
\bottomrule
\end{tabular}
\end{table}

Table~\ref{tab:constants} lists the results, and Figure~\ref{fig:constants} plots them on a logarithmic scale grouped by frequency. The two per-report operations, HMAC and Merkle hashing, cost 2.15 and 0.66\,$\mu$s. The per-epoch ML-KEM-768 decapsulation costs 95.6\,$\mu$s. The per-checkpoint operations span three orders of magnitude: an ML-DSA-65 signature costs 254.9\,$\mu$s, a 4\,KiB \texttt{write}+\texttt{fsync} costs 4.24\,ms, and an SLH-DSA-128f signature costs 31.52\,ms. The signature cost depends on the library. On the same host, liboqs signs ML-DSA-65 in 0.255\,ms and OpenSSL~3.5.6 in 1.502\,ms.

The durable write costs more than sixteen ML-DSA-65 signatures and therefore dominates checkpoint cost under ML-DSA. More broadly, checkpoint operations cost two to four orders of magnitude more than per-report operations. The next experiment measures how checkpoint interval amortizes this cost and impacts the detection delay.

\subsection{Checkpoint Amortization and Omission Detection}\label{sec:model}
RQ1 asks whether checkpointing reduces repeated post-quantum cost enough to justify the delay it introduces before an omission becomes visible. The preceding experiment provides the protocol-operation costs. This experiment varies the checkpoint interval and measures how those costs amortize across reports as the omission-detection delay increases.

Let $C_{\mathrm{mac}}$ and $C_{\mathrm{leaf}}$ denote the per-record MAC and Merkle costs, and let $C_{\mathrm{sign}}$ and $C_{\mathrm{fsync}}$ denote the checkpoint signature and durable-write costs. For an interval of $k$ reports, the amortized controller cost per report is
\begin{equation}\label{eq:model}
\mathrm{cost}(k)=C_{\mathrm{mac}}+C_{\mathrm{leaf}}+
\frac{C_{\mathrm{sign}}+C_{\mathrm{fsync}}}{k}.
\end{equation}
The first two terms form a floor of 2.81\,$\mu$s per report. The amortized checkpoint cost equals that floor at
\[
k^\star=
\frac{C_{\mathrm{sign}}+C_{\mathrm{fsync}}}
{C_{\mathrm{mac}}+C_{\mathrm{leaf}}}.
\]
The same interval bounds detection delay. At an aggregate report rate $r$, an omission remains hidden for at most $W=k/r$.

We swept \(k\) from 1 to 4096 on \(C\) at 2400,MHz, using five trials per point and three baselines: ML-DSA-65 without a durable log, ML-DSA-65 with a durable log, and SLH-DSA-128f with a durable log. These baselines separate signature amortization from durable-write amortization. Because ML-DSA and SLH-DSA are stateless, they require a durable write only when the deployment preserves the evidence log across restart. In contrast, stateful LMS, HSS, and XMSS require a durable write for every signature. To test the detection bound, we inserted one omission at a random position in each batch and ran 200 trials per interval. To test equivocation, we ran 200 trials in which the controller signed conflicting checkpoints for the same position.

\begin{table}[t]
\caption{Amortized cost per report at $k=1$ and $k=4096$. Controller $C$ at 2400\,MHz; five trials per point.}
\label{tab:baselines}
\centering
\footnotesize
\setlength{\tabcolsep}{3pt}
\begin{tabular}{@{}lcrrr@{}}
\toprule
Signer & Durable log & $k=1$ & $k=4096$ & Reduction \\
\midrule
ML-DSA-65    & No  & 306.2\,$\mu$s & 3.40\,$\mu$s  & $90\times$ \\
ML-DSA-65    & Yes & 4354\,$\mu$s  & 4.96\,$\mu$s  & $878\times$ \\
SLH-DSA-128f & Yes & 36.2\,ms      & 12.15\,$\mu$s & $2979\times$ \\
\bottomrule
\end{tabular}
\end{table}

\begin{figure}[t]
\centering
\includegraphics[width=\columnwidth]{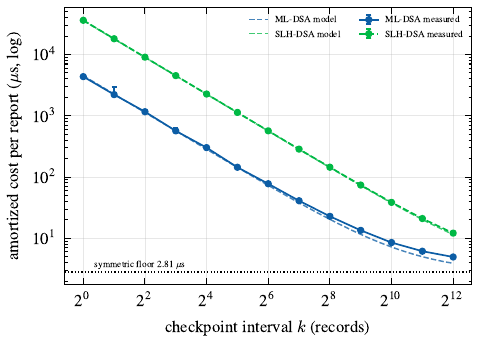}
\caption{Measured amortized cost against Equation~\ref{eq:model}. The dotted line marks the
2.81\,$\mu$s per-report floor. Controller $C$ at 2400\,MHz.}
\label{fig:cost}
\end{figure}

\begin{figure}[t]
\centering
\includegraphics[width=\columnwidth]{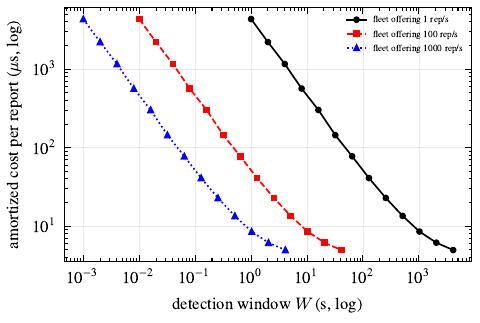}
\caption{RQ1 cost-detection tradeoff. At aggregate report rate $r$, a checkpoint interval of $k$ reports gives a maximum omission-detection delay of $W=k/r$.}
\label{fig:pareto}
\end{figure}

Figure~\ref{fig:cost} overlays the model on the sweep. The median relative error is 1.2\,\% for SLH-DSA and 6.3\,\% for ML-DSA. Table~\ref{tab:baselines} gives the endpoints. Checkpointing reduces the ML-DSA-65 cost per report by $90\times$ without a durable log and by $878\times$ with one, and it reduces the SLH-DSA-128f cost by $2979\times$. The amortizable cost has a different makeup under each scheme. Under ML-DSA-65, the durable write accounts for 94.3\,\% of it and the signature for 5.7\,\%, which puts $k^\star$ at 1602. Under SLH-DSA-128f, the signature accounts for 88.1\,\% and the write for 11.9\,\%, which puts $k^\star$ at 12739. In the omission trials, the next checkpoint exposed the omission after a median of about $k/2$ records and never after more than $k$: 36 and 64 records at $k=64$, and 2014 and 4079 records at $k=4096$. The range-width check caught every injected omission. In the equivocation trials, comparing the two signed views exposed the conflict in all 200 cases.

These results answer RQ1. Increasing $k$ reduces per-report cost by amortizing checkpoint operations over more evidence, but proportionally increases the maximum delay before a checkpoint-visible omission can be exposed. No single checkpoint interval is optimal across deployments; the appropriate value depends on the acceptable detection window and the cost structure of the selected signature scheme. The experiments in this section maintains a fixed device population, so the next experiment asks whether the amortized cost survives as the device count grows.

\subsection{Scaling Across Controller Deployments}\label{sec:fleet}
While deploying a series of controllers each device adds a sequence range and subtree root to every checkpoint. This experiment measures how controller cost per report, checkpoint size, and memory use change as the number of device identities grows.

We assigned one report to each device per checkpoint, so the number of identities and the checkpoint batch size increased together. At 10, 100, 1,000, 4,000, and 10,000 identities, we measured controller cost per report, checkpoint size, checkpoint verification time, and resident memory using ML-DSA-65 checkpoints and five trials per point.

\begin{table}[t]
\caption{Controller-side device scaling with ML-DSA-65 checkpoints; five trials per point.}
\label{tab:fleet}
\centering
\begin{tabular}{@{}rrrrrr@{}}
\toprule
Devices & $\mu$s/report & Checkpoint & B/report & Verify & RSS \\
\midrule
10 & 491.0 & 3.9\,kB & 495.0 & 0.117\,ms & 6.6\,MB \\
100 & 51.3 & 8.8\,kB & 189.0 & 0.134\,ms & 6.6\,MB \\
1\,000 & 10.84 & 58.0\,kB & 158.4 & 0.302\,ms & 6.6\,MB \\
4\,000 & 6.93 & 222.1\,kB & 155.9 & 0.866\,ms & 7.6\,MB \\
10\,000 & 6.18 & 550.2\,kB & 155.3 & 1.986\,ms & 8.3\,MB \\
\bottomrule
\end{tabular}
\end{table}
Table~\ref{tab:fleet} shows that amortized controller cost falls from 491.0\,$\mu$s per report at 10 identities to 6.18\,$\mu$s at 10\,000 because fixed checkpoint work is spread across more records. Checkpoint size reaches 550.2\,kB at 10\,000 identities, while overhead converges to 155.3 octets per report. Checkpoint verification takes 1.986\,ms at 10\,000 identities. The implementation stores 56 octets of policy state per identity and reaches 8.3\,MB RSS.

Controller CPU and memory remain within the measured platform's capacity at 10\,000 identities, but checkpoint size grows with fleet size because each identity contributes metadata. 

\subsection{Cost of Transferable Device-Origin Authentication}\label{sec:twotag}
Checkpointing preserves inclusion and order, but a shared evidence MAC does not provide transferable device-origin authentication. RQ2 asks whether the disclosure chain restores that property without returning public-key work to every report.

\begin{figure}[t]
\centering
\includegraphics[width=\columnwidth]{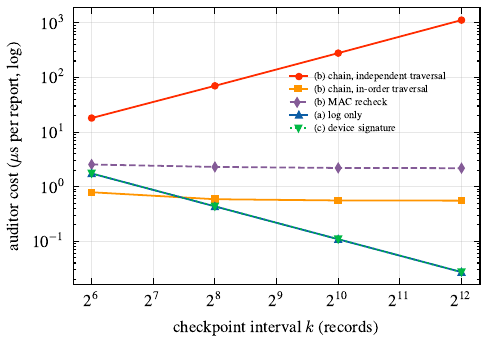}
\caption{Cold-auditor verification cost per report. Independent traversal hashes each disclosed key back to the chain anchor, requiring $O(j)$ work, while in-order traversal advances from the last verified key in $O(1)$ work.}
\label{fig:audit}
\vspace{-.3cm}
\end{figure}

We evaluated the two-tag design in three nodes. First, on $C$ at 10\,000 identities, we compared one-tag and two-tag reports. We alternated the two configurations within each trial so that both experienced similar thermal and cache conditions. Next, on $P$, we measured the cost of generating the second MAC and advancing the disclosure chain. Finally, at the auditor, we measured two verification methods. Independent verification hashes each disclosed $K_j$ back to the anchor. In-order verification keeps the last verified key and advances the chain by one step. Figure~\ref{fig:audit} compares the three audit levels from \S\ref{sec:audit} across values of $k$.

Adding the second tag had negligible controller cost. The prover spent 5.54\,$\mu$s per report on the second MAC and disclosure-chain step. Auditor cost depended on traversal order: independent traversal increased with $k$, while in-order traversal remained below 1\,$\mu$s per report. Including the MAC recheck, in-order device-origin verification cost 2.95\,$\mu$s per report.

These results answer RQ2. AMORT-V1 restores transferable device-origin authentication without a public-key operation on each report. In-order traversal keeps verification in the microsecond range, while key disclosure adds one checkpoint interval of delay. The next experiment tests whether the added prover work remains within the PLC control-cycle budget.

\subsection{Impact on PLC Control-Cycle Timing}\label{sec:realtime}

We measure scan response time from scheduled release to completion over $3\times10^{4}$ cycles in each of seven configurations on prover $P$ at 1500,MHz to test whether one- or two-tag evidence generation at 1,Hz causes deadline misses. We also test whether background checkpoint signing on the same core affects the scan deadline.
 The configurations include a baseline, one and two-tag evidence generation at both 1\,Hz and every control cycle, and background ML-DSA-65 and SLH-DSA-128f signing co-located on the same core.

\begin{figure}[t]
\centering
\includegraphics[width=\columnwidth]{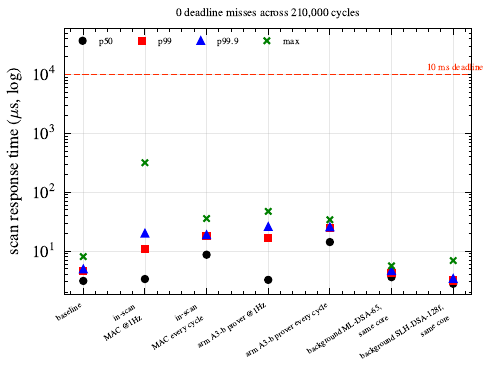}
\caption{Scan response-time distribution relative to the 10ms deadline}
\label{fig:realtime}
\vspace{-.3cm}
\end{figure}

Figure~\ref{fig:realtime} shows no deadline misses across $2.1\times10^{5}$ total cycles. Median response time remains between 2.79 and 14.27\,$\mu$s against a 10\,ms control deadline, and the largest observed response is 317\,$\mu$s. Generating one MAC per scan increases median execution time from 0.130 to 5.444\,$\mu$s, while generating two tags together with the disclosure-chain step increases it to 10.981\,$\mu$s. Evidence generation and background signing therefore leave more than an order of magnitude of headroom under the 10\,ms deadline.

\subsection{Failure Isolation and Rollback-Safe Recovery}\label{sec:recovery}
RQ3 asks whether per-device epochs contain failures and support recovery without returning durable writes to each report. We separate the question into three failure modes: revocation and re-keying, loss of uncommitted evidence during restart, and rollback of sequence state.

For isolation, we revoke one identity in deployments of 100, 1\,000, and 10\,000 identities and measure whether other identities continue attesting. We also measure re-keying cost per identity. For partition recovery, we replay buffered evidence for backlogs from 64 to 4096 records. For restart, we terminate a physical prover at controller-unknown points with $k=64$ and compare the last seen and last committed sequence numbers in \textsc{EpochClose}. For rollback, we compare fresh-epoch recovery, an unanchored reservation scheme, and a TPM-backed reservation scheme at $R=64$. The rollback study runs 30 trials per cell on each of five provers, for 150 trials per arm.

Revoking one identity leaves all other identities attesting at 100, 1\,000, and 10\,000 identities. Re-keying costs 35.1--35.4\,$\mu$s per identity: 3.54\,ms for 100 identities, 35.2\,ms for 1\,000, and 351\,ms for 10\,000. Re-merging buffered partition evidence costs 2.35\,$\mu$s per record for backlogs from 64 to 4096 records. At a 1\,Hz report rate, replaying a one-hour backlog requires less than 9\,ms of controller time.

In the restart test, two interrupted epochs lose 22 and 34 uncommitted records and raise \texttt{ERR\_EVIDENCE\_INCOMPLETE}. A clean epoch loses no records and raises no error. The loss remains bounded by the checkpoint interval.

Table~\ref{tab:tpm} outlines rollback behavior. Across 150 rollback trials, the unanchored scheme re-issues 19\,200 sequence numbers. The TPM-backed scheme detects all 150 stale states and re-issues none. TPM-backed restart reaches ready state in 10.39\,ms, compared with 5.63\,ms for fresh-epoch recovery. Crash injection after the counter increment fails secure; crashes within a valid reserved window or after durable reservation resume without sequence reuse. A complete TPM-backed reservation costs 15.77\,ms. With an assumed NV-write budget of $5\times10^5$ writes and one report per second, a ten-year service life requires $R\geq631$.

These experiments answer RQ3. Epoch state confines revocation and re-keying to one identity, fresh-epoch recovery makes any lost evidence visible, and TPM-backed reservation stops sequence reuse under rollback while keeping the TPM off the per-report path.

\begin{table}[t]
\caption{Restart and rollback behavior at $R=64$; 30 trials per cell on five provers.}
\label{tab:tpm}
\centering
\begin{tabular}{@{}llrrr@{}}
\toprule
& Arm & New epoch & Ready (ms) & Seq.\ reuse \\
\midrule
\multirow{3}{*}{Restart}
 & fresh  & 150/150 & 5.63  & 0 \\
 & naive  & 0/150   & 1.40  & 0 \\
 & tpm    & 0/150   & 10.39 & 0 \\
\midrule
\multirow{3}{*}{Rollback}
 & fresh  & 150/150 & --- & 0 \\
 & naive  & 0/150   & --- & 19\,200 \\
 & tpm    & 150/150 & --- & 0 \\
\bottomrule
\end{tabular}
\vspace{-.1cm}
\end{table}

\subsection{Full-Image Measurement as an Attestation Bottleneck}
\label{sec:image}
The preceding experiments use a 4\,KiB cached representation of software state to isolate authentication and checkpoint costs. That setup does not show whether checkpointing remains beneficial when each attestation recomputes the full software image. We therefore measure the cost of generating the underlying OpenPLC measurement.

On each of the five physical provers, we measure an OpenPLC~v3 installation containing 6\,239 files and 355\,MB. The prover hashes every component, sorts the $(\mathrm{path},\mathrm{digest})$ pairs, and hashes the resulting list. We run five repetitions per host under warm and cold-cache conditions. We also compare the resulting image digests across the five hosts to identify reference-value instability.

A warm-cache image measurement takes 392\,ms, or $1\,511\times$ of one ML-DSA-65 signature, and supports about 2.6 measurements per second. A cold-cache measurement takes 8.352\,s, or $32\,752\times$ of one ML-DSA-65 signature, and supports about 0.12 measurements per second.

Checkpointing reduces authentication and persistence cost, but it does not reduce measurement cost. Equation~\ref{eq:model} therefore applies when the deployment avoids a full image read for every report through caching, incremental measurement, hardware support, or another state-maintenance mechanism.

\begin{table}[t]
\caption{Full-image measurement on five physical provers at 2400\,MHz; five repetitions per host.}
\label{tab:image}
\centering
\begin{tabular}{@{}lrrr@{}}
\toprule
 & Median & Throughput & vs ML-DSA-65 sign \\
\midrule
Warm cache & 392\,ms & 921\,MB/s & $1\,511\times$ \\
Cold cache & 8\,352\,ms & 42.5\,MB/s & $32\,752\times$ \\
\bottomrule
\end{tabular}
\vspace{-.4cm}
\end{table}

\section{Conclusion}
\vspace{-.1cm}
AMORT-V1 shows that post-quantum attestation does not require a public-key signature and durable write for every report. By moving these operations to periodic checkpoints, the protocol separates the fast path for continuous evidence generation from the slower path for public verification and durable commitment. This structure turns repeated post-quantum cost into a configurable tradeoff between attestation overhead and detection delay.

Checkpointing alone is not sufficient because symmetric evidence cannot prove device origin to a third party. AMORT-V1 closes this gap with a device-held disclosure chain that restores transferable device-origin authentication after commitment. These mechanisms provide a path toward long-lived post-quantum attestation for controller fleets without placing expensive cryptographic or durable-state operations on every report.

We also show checkpointing can amortize authentication and persistence, but it cannot reduce the cost of repeatedly measuring a large software image. Future work should therefore combine AMORT-V1 with incremental measurement, hardware-assisted measurement, and trusted measurement-state caching. It should also study checkpoint policies that adapt to report rate, device criticality, and acceptable detection delay, and evaluate the protocol on larger heterogeneous PLC fleets and production control networks. These directions would move checkpointed attestation from a fixed protocol design toward an adaptive post-quantum trust architecture for industrial systems.

\appendices
\newpage
\section{The Attack Matrix}\label{app:attacks}

Table~\ref{tab:attack} lists the 27 adversarial cases and five positive controls that validate the reference model. The cases cover cryptographic tampering, decoder errors, and protocol-semantic violations, including replay, stale epochs, omission, equivocation, early disclosure, tag-length downgrade, restart replay, and malformed inputs.
\vspace{-5cm}

\begin{table*}[t]
\centering
\caption{27-case attack matrix evaluated against the AMORT-V1 reference model \\ Class: \textbf{C} cryptographic,
\textbf{D} decoder, \textbf{S} protocol-semantic.}
\label{tab:attack}
\footnotesize
\setlength{\tabcolsep}{4pt}
\begin{tabular}{@{}llp{2.9cm}l@{\hspace{0mm}}llp{2.9cm}l@{}}
\toprule
\# & Cl. & Adversary action & Rejected by & \# & Cl. & Adversary action & Rejected by \\
\midrule
A01 & C & Measurement modified after MAC & \textsc{mac\_fail} & A15 & S & Controller equivocates between verifiers & \textsc{ckpt\_fork} \\
A02 & C & Sequence number modified after MAC & \textsc{mac\_fail} & A16 & S & Checkpoint withheld from the chain & \textsc{ckpt\_discontinuity} \\
A03 & C & Device identifier modified after MAC & \textsc{mac\_fail} & A17 & S & Chain key disclosed before commitment & \textsc{disclose\_early} \\
A04 & C & MAC tag truncated & \textsc{truncated} & A18 & S & MAC tag length downgraded & \textsc{taglen\_downgrade} \\
A05 & C & Checkpoint root modified after signing & \textsc{sig\_fail} & A19 & S & Whole batch suppressed for one device & \textsc{seq\_gap} \\
A06 & C & Disclosed key not on the committed chain & \textsc{chain\_fail} & A20 & S & Controller omits and substitutes a record & \textsc{proof\_fail} \\
A07 & C & Forged Merkle inclusion proof & \textsc{proof\_fail} & A21 & S & Controller forges record and discloses & \textsc{mac\_fail} \\
A08 & C & Valid record replayed under another device & \textsc{mac\_fail} & A22 & C & Controller state exposes chain material & \textsc{chain\_fail} \\
A09 & S & Correctly MACed record replayed & \textsc{seq\_duplicate} & A23 & S & Device replays sequence after restart & \textsc{mac\_fail} \\
A10 & S & Device skips a sequence number & \textsc{seq\_gap} & A24 & D & Record truncated & \textsc{truncated} \\
A11 & S & Record from a superseded epoch & \textsc{epoch\_stale} & A25 & D & Trailing bytes appended & \textsc{trailing} \\
A12 & S & Evidence from an unregistered device & \textsc{device\_unknown} & A26 & D & Domain-separation tag corrupted & \textsc{bad\_domain} \\
A13 & S & Evidence from a revoked device & \textsc{revoked} & A27 & D & Unknown protocol version & \textsc{bad\_version} \\
A14 & S & Controller omits a record mid-batch & \textsc{seq\_gap} &  &  &  &  \\
\midrule
\multicolumn{8}{@{}l}{\textbf{Positive controls} --- the honest path must still succeed} \\
P1 &  & Honest stream accepted by a cold auditor & 2 checkpoints verified & P2 &  & Inclusion proof verifies with no secret & path length 4 \\
P3 &  & Public origin authentication after disclosure & arm A3-b, 8 records & P4 &  & Revoking one device leaves the fleet attesting & peer unaffected \\
P5 &  & One durable operation per checkpoint & 640 records, 10 fsync at $k$=64 &  &  &  &  \\
\bottomrule
\end{tabular}
\\[2pt]
\footnotesize All 27 cases reject with the expected code and all 5 controls pass (9 crypto, 4 decoder, 14 semantic).
\end{table*}

\end{document}